\documentclass[aps,prl,twocolumn,superscriptaddress,showpacs,amsmath,amssymb]{revtex4-1}

\pdfoutput=1
\usepackage{graphicx}
\usepackage{dcolumn}
\usepackage{bm}

\newcommand{\mr}[1]{\mathrm{#1}}
\newcommand{\be}{\begin{equation}}
\newcommand{\ee}{\end{equation}}

\newcommand{\kb}{k_{\mr{B}}}

\newcommand{\dds}{d_{\mr{S}}}
\newcommand{\ddn}{d_{\mr{N}}}

\newcommand{\lo}{\mathcal{L}_0}

\newcommand{\figta}{$\left(\mathrm{a}\right)\;$}
\newcommand{\figtb}{$\left(\mathrm{b}\right)\;$}
\newcommand{\figtc}{$\left(\mathrm{c}\right)\;$}
\newcommand{\figtd}{$\left(\mathrm{d}\right)\;$}
\newcommand{\figte}{$\left(\mathrm{e}\right)\;$}

\newcommand{\figa}{$\left(\mathrm{a}\right)$}
\newcommand{\figb}{$\left(\mathrm{b}\right)$}
\newcommand{\figc}{$\left(\mathrm{c}\right)$}
\newcommand{\figd}{$\left(\mathrm{d}\right)$}
\newcommand{\fige}{$\left(\mathrm{e}\right)$}

\newcommand{\equref}[1]{(\ref{#1})}
\newcommand{\sigman}{\sigma_{\mr{N}}}

\newcommand{\tbath}{T_0}
\newcommand{\vcool}{V}

\newcommand{\vth}{V_{\mr{th}}}
\newcommand{\ith}{I_{\mr{th}}}

\newcommand{\rcool}{R_{\mr{T}}}

\newcommand{\voln}{\mathcal{V}_{\mr{N}}}
\newcommand{\an}{\mathcal{A}_{\mr{N}}}

\newcommand{\tn}{T_{\mr{N}}}
\newcommand{\ts}{T_{\mr{S}}}
\newcommand{\qdotn}{\dot{Q}_{\mr{N}}}
\newcommand{\qdots}{\dot{Q}_{\mr{S}}}
\newcommand{\ns}{n_{\mr{S}}}
\newcommand{\vopt}{V_{\mr{opt}}}
\newcommand{\tph}{T_{\mr{ph}}}
\newcommand{\tnmin}{T_{\mr{N,min}}}
\newcommand{\tno}{T_{\mr{N},0}}

\newcommand{\tc}{T_{\mr{C}}}

\newcommand{\nno}{N_{0}}
\newcommand{\po}{P_{0}}

\newcommand{\sigmaephn}{\Sigma_{\mr{N}}}
\newcommand{\kappas}{\kappa_{\mr{S}}}
\newcommand{\kappan}{\kappa_{\mr{N}}}

\newcommand{\across}{A}
\newcommand{\bperp}{B_{\perp}}
\newcommand{\deltat}{\delta T}
\newcommand{\bbo}{B_0}

\newcommand{\kohm}{\;\mr{k}\Omega}

\newcommand{\gauss}{\;\mr{G}}
\newcommand{\mk}{\;\mr{mK}}
\newcommand{\wb}{\;\mr{Wb}}
\newcommand{\fw}{\;\mr{fW}}

\newcommand{\muohmcm}{\;\mu\Omega\mr{cm}}

\newcommand{\mum}{\;\mu\mr{m}}
\newcommand{\mut}{\;\mu\mr{T}}
\newcommand{\muev}{\;\mu\mr{eV}}
\newcommand{\nm}{\;\mr{nm}}
\newcommand{\kelvin}{\;\mr{K}}

\begin{document}

\title{Magnetic-Field-Induced Stabilization of Non-Equilibrium Superconductivity}

\author{J. T. Peltonen}
\affiliation{Low Temperature Laboratory, Aalto University, P.O. Box 13500, FI-00076 AALTO, Finland}

\author{J. T. Muhonen}
\affiliation{Low Temperature Laboratory, Aalto University, P.O. Box 13500, FI-00076 AALTO, Finland} \affiliation{Department of Physics, University of Warwick, CV4 7AL, UK}

\author{M. Meschke}
\affiliation{Low Temperature Laboratory, Aalto University, P.O. Box 13500, FI-00076 AALTO, Finland}

\author{N. B. Kopnin}
\affiliation{Low Temperature Laboratory, Aalto University, P.O. Box 13500, FI-00076 AALTO, Finland}
\affiliation{ L.~D.~Landau
Institute for Theoretical Physics, 117940 Moscow, Russia}

\author{J. P. Pekola}
\affiliation{Low Temperature Laboratory, Aalto University, P.O. Box 13500, FI-00076 AALTO, Finland}

\date{\today}

\begin{abstract}
A small magnetic field is found to enhance relaxation processes in a superconductor thus \emph{stabilizing superconductivity} in non-equilibrium conditions. In a normal-metal (N) -- insulator -- superconductor (S) tunnel junction, applying a field of the order of $100\mut$ leads to significantly \emph{improved} cooling of the N island by quasiparticle (QP) tunneling. These findings are attributed to faster QP relaxation within the S electrodes as a result of enhanced QP drain through regions with locally suppressed energy gap due to magnetic vortices in the S leads at some distance from the junction.
\end{abstract}


\maketitle

\begin{figure}[!htb]
\includegraphics[width=\columnwidth]{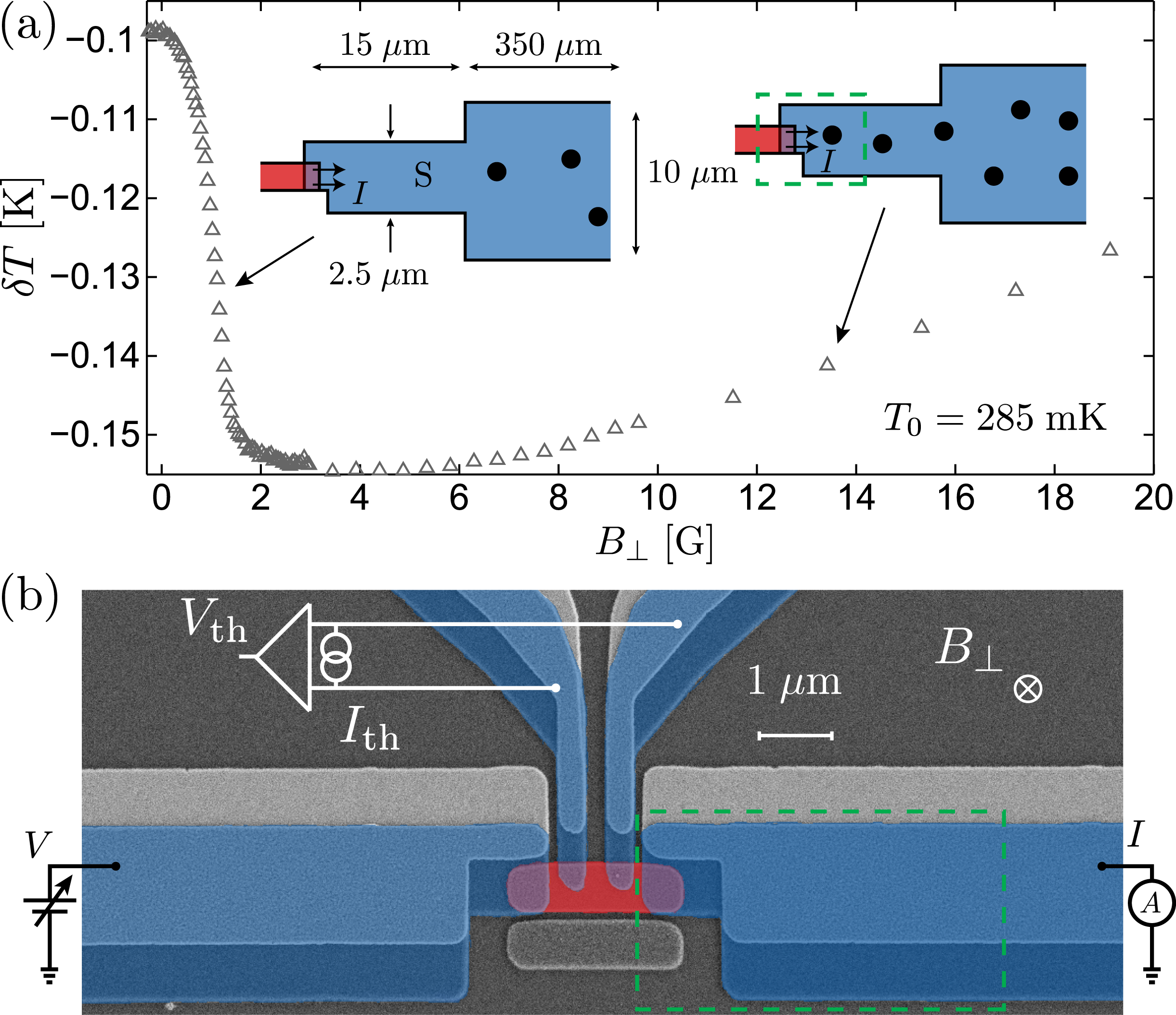}
\caption{(color online) \figta Maximum temperature drop $\deltat$ in an optimally biased SINIS cooler in a perpendicular magnetic field $\bperp$, at the bath temperature $\tbath=285\mk$. The sketches show the S electrode geometry and qualitative vortex configurations at $\bperp\lesssim 2\gauss$ and at a value of $\bperp$ beyond the optimum point. The area inside the green dashed rectangle corresponds to that in the micrograph below. \figtb Scanning electron micrograph of a typical structure, together with the measurement scheme (see text for details). A Cu island (red) is contacted to four superconducting Al electrodes (blue) via Al oxide tunnel barriers for thermometry and temperature control. Replicas of each structure are visible due to the fabrication involving two-angle shadow evaporation of the metals.} \label{fig:scheme}
\end{figure}

In this Letter, we report an observation that appears counterintuitive at first: a small magnetic field is found to stabilize superconductivity under quasiparticle (QP) injection. In our experiment, the cooling power of normal-metal (N) -- insulator (I) -- superconductor (S) tunnel structures is enhanced in perpendicular magnetic fields $\bperp\simeq 100\mut=1\gauss$. Measured maximum temperature drop $\deltat$ relative to a starting bath temperature $\tbath=285\mk$ exhibiting this behavior is shown in Fig.~\ref{fig:scheme}~\figa. The improvement is unexpected, as in general the effect of a magnetic field is to suppress superconductivity. Electronic cooling in NIS junctions in the presence of magnetic fields in both perpendicular and parallel orientations has been studied before~\cite{arutyunov00}, but only in higher fields where the cooling power was already reduced due to diminishing superconducting energy gap $\Delta$. On the other hand, the creation of magnetic vortices~\cite{abrikosov57} has been shown to enhance QP relaxation in superconducting aluminum, as the QPs become trapped and thermalize in the regions of reduced $\Delta$~\cite{ullom98b}. Here we demonstrate that the additional relaxation channel due to enhanced QP drain through regions occupied by magnetic vortices enhances the superconducting performance of S leads and improves the electronic cooling in NIS junctions. This can be of relevance in superconducting qubits~\cite{martinis09,catelani11,lenander11,gunnarsson04}, resonators~\cite{devisser11} and in hybrid SINIS turnstiles~\cite{pekola08} in reducing the effects from nonequilibrium and residual QPs arising due to drive and microwave radiation from the environment. Moreover, improved relaxation caused by vortex creation in the S leads can partially explain the ``re-entrant superconductivity'' observed in Zn and Al nanowires~\cite{chen09,chen11}. In the present case, as sketched in Fig.~\ref{fig:scheme}~\figa, vortex formation in the S electrodes away from the NIS junction improves relaxation of the injected QPs and leads to enhanced cooling of the N island. In higher fields, vortices move closer to the junction, deteriorating the cooling power.

In an NIS junction, $\Delta$ acts as an energy filter for the tunneling QP~\cite{nahum93,nahum94,giazotto06,leivo96}. At low temperatures $\kb T\ll \Delta$ and for bias voltages $eV\lesssim\Delta$ across the junction, the electrons in the N electrode cool considerably below the phonon temperature by hot QP extraction. The effect can be made more pronounced in a symmetric double junction SINIS structure with a small N island contacted to S leads via two NIS junctions~\cite{leivo96}, allowing to construct practical solid-state refrigerators for cooling thin-film detectors to temperatures close to $100\mk$~\cite{clark04,clark05}. The performance of actual devices depends crucially on the relaxation of the QPs that are injected into the S electrode, as the superconductor overheating diminishes the cooling power at an NIS junction because of enhanced QP backtunneling. The excess QP density close to the junction can be diminished by fabricating the S electrodes very thick~\cite{clark04}, or covering them partially by a layer of normal metal that acts as a QP trap~\cite{joyez94,pekola00,court08}. The QP population is typically modeled in terms of a diffusion equation, describing their recombination retarded by phonon retrapping, and other loss mechanisms~\cite{rothwarf67,kaplan76,ullom98a,ullom00a,rajauria09}. Converting the excess density into an effective, position-dependent temperature $T(x)$~\cite{ullom00b,arutyunov11}, one finds that at phonon temperatures $\kb T\ll \Delta$ the S leads can be overheated on a length scale ranging from tens of micrometers to a millimeter, as the electron-phonon relaxation and electronic heat conduction are exponentially suppressed compared to their normal state values~\cite{kaplan76,bardeen59}.

Here we present data from one of several measured symmetric SINIS structures similar to that shown in Fig.~\ref{fig:scheme}~\figb, fabricated at different times and electrically characterized in a dilution refrigerator down to $50\;\mk$ bath temperature. The same qualitative behavior was observed in all structures with the same geometry. A copper island of area $\an\simeq 2.7\times 0.7\mum^2$ is contacted by four overlap-type Al/Al-oxide/Cu NIS junctions. The Al electrodes with zero-temperature energy gap $\Delta_0\simeq 210\muev$ become superconducting below $\tc\simeq 1.4\kelvin$. Compared to the two small (probe) junctions in the middle, the two outer (cooler) junctions at the ends of the island have larger overlap area and therefore lower normal state tunnel resistance $\rcool\simeq 1.1\kohm$ each. The structures were fabricated on an oxidized silicon substrate by standard electron beam lithography and two-angle shadow evaporation of Al and Cu through a polymer resist mask. First, an Al layer of thickness $\dds\simeq 30\nm$ was deposited, followed by \emph{in situ} oxidation in the e-beam evaporator chamber in a few millibars of pure oxygen for a few minutes. Finally, a Cu layer of thickness $\ddn\simeq 30\nm$ was evaporated at a different angle, forming the N island with four tunnel contacts to the Al electrodes. In addition, Cu replicas of the Al leads form large area tunnel junctions by partly covering the Al layer, serving as QP traps, albeit of suboptimal performance~\cite{pekola00}.

\begin{figure}[!htb]
\includegraphics[width=\columnwidth]{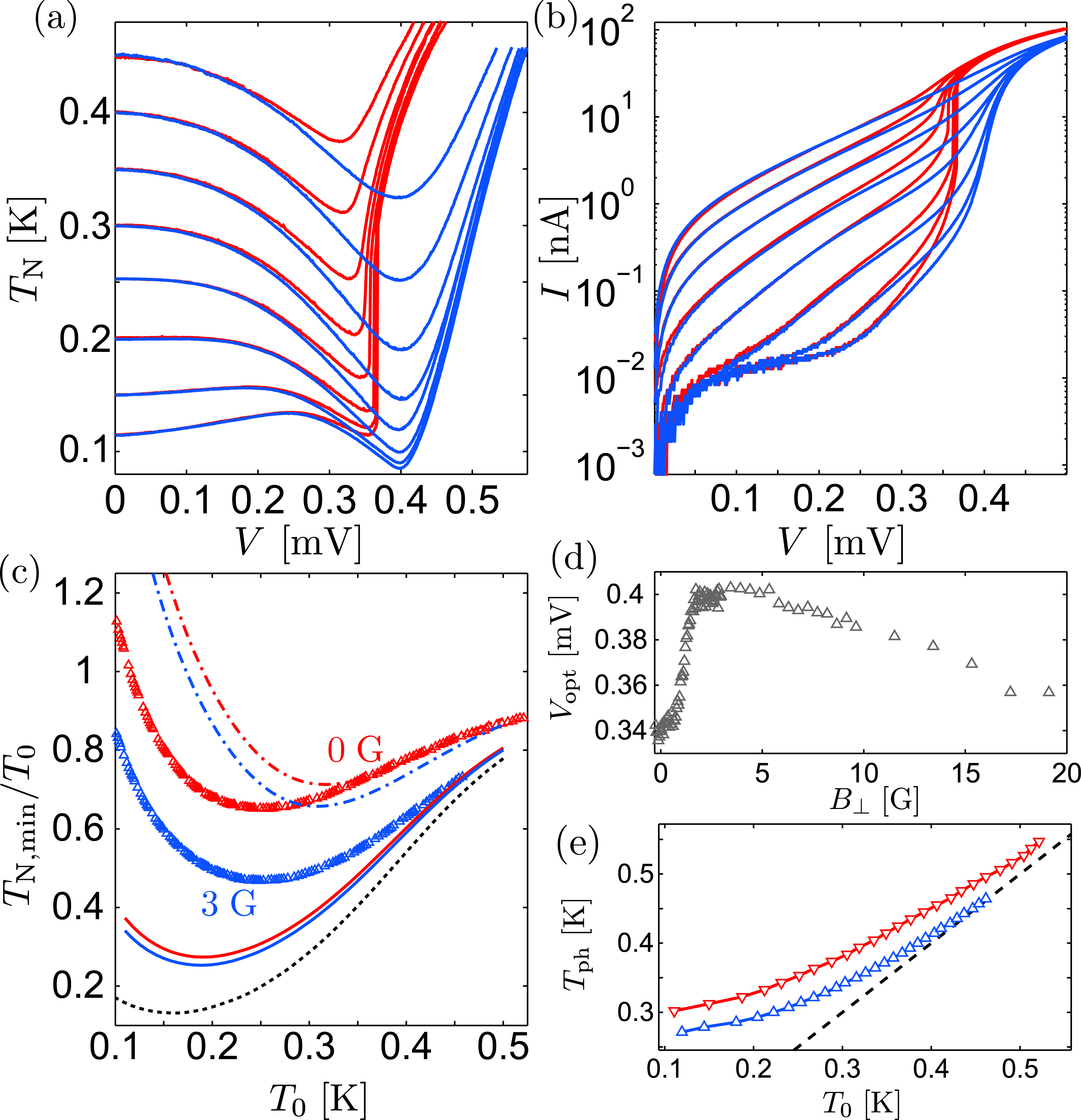}
\caption{(color online) \figta Temperature $\tn$ of the N island and \figtb cooler SINIS IV characteristic at several bath temperatures $\tbath$ as functions of the cooler bias voltage $\vcool$, in zero field (red) and $\bperp=3\gauss$ (blue). \figtc Relative minimum temperature in zero field (red symbols) and at $\bperp\simeq3\gauss$ (blue symbols). The solid and dashed-dotted lines show the calculated temperature reduction for various degrees of thermalization of the QPs and of the island phonons (see text).  \figtd Optimum bias voltage $\vopt$ {\it vs} $\bperp$ at $\tbath=285\mk$, with the corresponding temperature drop in Fig.~\ref{fig:scheme}~\figa. \figte Phonon temperatures in zero (red) and optimum (blue) field, required to reproduce the observed $\tnmin$. Dashed line shows $\tph=\tbath$.} \label{fig:zerosmall}
\end{figure}

The island electron temperature $\tn$ is measured by biasing the probe junctions by a constant current $\ith$, and measuring the voltage drop $\vth$, calibrated against $\tbath$ at $\vcool=0$. The solid lines in Fig.~\ref{fig:zerosmall}~\figta show the measured $\tn$ as a function of $\vcool$ at bath temperatures $\tbath$ between 0.1 and $0.5\kelvin$, in zero field (red) and at $\bperp=3\gauss$ (blue). In the following, the minimum $\tn$ along each curve at a fixed $\tbath$ is denoted by $\tnmin$, and the corresponding bias voltage by $\vopt$. The strong influence of small $\bperp$ on the cooling is evident: At $e\vcool\simeq 2\Delta$ the maximum cooling $\deltat=\tnmin-\tno$ at each $\tbath$ increases by several tens of percents. The cooler QP current displays analogous behavior, Fig.~\ref{fig:zerosmall}~\figb. At the same time the optimum bias voltage $\vopt$ increases, Fig.~\ref{fig:zerosmall}~\figd, while heating at $\vcool>\vopt$ diminishes. The cooling enhancement is symmetric in the applied field.

The improved refrigeration is summarized in Fig.~\ref{fig:zerosmall}~\figc, where the symbols show the $\tbath$-dependent relative minimum temperature in zero field and close to optimum $\bperp$.
We observed the improved cooling also with Ag as the normal metal, in single NIS junctions with various gradually widening lead geometries close to the junction, and in a parallel field. In the latter case, the required fields were larger by an order of magnitude and dependent on the field orientation in the sample plane.

The thin-film Al leads behave as a type II superconductor, so that $\bperp$ penetrates in the form of vortices~\cite{tinkhambook,tinkhamharper}. Based on a typical normal state resistivity $\rho=3.5\muohmcm$ of our Al at $4.2\kelvin$~\cite{timofeevpeltonen}, the elastic mean free path is $l\simeq 8\nm$. With the Bardeen-Cooper-Schrieffer (BCS) coherence length $\xi_0\simeq 1600\nm$ and the London penetration depth $\lambda_{\mr{L}}\simeq 16\nm$ for bulk pure Al at low-temperatures, one obtains $\xi =0.855(\xi_0 l)^{1/2}\simeq 100\nm$ and $\lambda=\lambda_{\mr{L}}(\xi_0/l)^{1/2}\simeq 230\nm$ for our Al films with the Ginzburg-Landau parameter $\kappa=\lambda/\xi \simeq 2.4>1/\sqrt{2}$ and the lower critical field for the bulk material $H_{c1}\lesssim 100\gauss$. As sketched in Fig.~\ref{fig:scheme}~\figa, the S leads of the cooler junctions have an initial width of approximately $1\mum$. At a distance of $1\mum$ away from the island, they widen to $2.5\mum$ width and continue for $15\mum$ before again widening to $10\mum$ width and connecting to large-area bonding pads further $350\mum$ away. The magnetic field below which vortices are completely expelled from a long and narrow S lead of width $W$ is of the order of $\bbo=\Phi_0/W^2$~\cite{stan04}, where $\Phi_0=h/(2e)\simeq 2\times 10^{-15}\wb$ is the magnetic flux quantum. For a strip of width $W=10\mum$, $\bbo\simeq 0.2\gauss$, whereas $W=2.5\mum$ results in $\bbo\simeq 3.3\gauss$. Taking into account the demagnetizing factor $1-n_z \sim (0.5 ... 2)\times 10^{-2}$ of our films we can conclude that the initial increase in $|\deltat|$ in Fig.~\ref{fig:scheme}~\figa ~observed below $1\gauss$ and the turn-back that starts close to $3\gauss$ are consistent with vortex penetration into the wide and narrow parts of the lead, respectively.

In Ref.~\onlinecite{ullom98b} with large area NIS junctions, the increased sub-gap conductance at small $\bperp$ could be directly associated with the fraction of vortices in the junction area. In contrast, we do not observe an increase in the cooler junction sub-gap current in the small fields. The thermometer junctions with narrower S electrodes are not considerably affected in fields $\bperp\lesssim 10\gauss$ even at bias voltages close to $2\Delta$, indicating that vortices exist only further away from these junctions.

We assume quasiequilibrium with a local electronic temperature  different from the bath temperature~\cite{giazotto06}. The cooling power depends on the temperatures $\tn$ and $\ts$ of N and S electrodes near the interface. They are found from the equation of heat balance in the N island
\be
2\qdotn(\vcool,\tn,\ts)=\sigmaephn\voln(T_{\rm ph}^5-\tn^5)+\po ,\label{nbalance}
\ee
and of the heat conduction in each superconducting lead,
\be
\nabla\cdot \left[\kappa \nabla T(x)\right]=q(x),\label{ts}
\ee
with the boundary conditions $\mp\kappa\across T'(0)=\qdots$, and $T(x)\rightarrow\tbath$ at $x\rightarrow\pm \infty$ for the right or left lead  ($\across$ is a wire cross section). The temperature of the lead at the interface is $\ts\equiv T(x=0)$. The first term in the rhs of Eq. (\ref{nbalance}) describes the heat transferred to phonons in the normal island. The island volume and electron-phonon coupling constant are $\voln=\an\ddn$ and $\sigmaephn$, respectively. $\po\simeq 1\fw$ in Eq.~(\ref{nbalance}) is a constant residual power due to imperfect RF-filtering of the measurement. The heat $\qdotn$ extracted from the island through a single NIS junction and the heat $\qdots$ injected into an S lead by tunneling are
\be
\dot{Q}_{\mr{N,S}}=\frac{1}{e^2\rcool}\! \int \! \ns(E_{\mr{S}})E_{\mr{N,S}}\left[f_{T_{\rm N}}(E_{\mr{N}})-f_{T_{\rm S}}(E_{\mr{S}})\right]  dE.\label{qdotns}
\ee
Here, $E_{\mr{N}}=E-e\vcool/2$, $E_{\mr{S}}=E$, $f_{T_{\mr{ N}},T_{\mr{ S}}}(E)=1/[\exp(E/\kb T_{\mr{N},\mr{S}})+1]$ are the Fermi occupation factors, and $\ns(E)=E/\sqrt{E^2-\Delta^2}]|$ is the normalized BCS density of states (DOS).

The rhs of Eq. (\ref{ts}) is the power transferred from the unit volume of the superconductor into the (unbiased) normal trap with temperature $\tbath$. Similarly to Eq. (\ref{qdotns}),
\begin{eqnarray*}
q(x)&=&\frac{1}{e^2 \rho_{\rm tr}d_{\rm S}}\int n_{\rm S}(E) E [f_{T}(E)-f_{T_0}(E)]\, dE \\
&=& [{\cal E}(T)-{\cal E}(T_0)]/\tau_{\rm tr}\ ,
\end{eqnarray*}
where $\tau_{\rm tr}^{-1}= 1/(2e^2 N_0 \rho_{\rm tr}d_{\rm S})$ is the time of relaxation to the trap, $\rho_{\rm tr}$ being the trap/superconductor tunnel resistance of unit contact area and $\nno$ denotes the normal state DOS at the Fermi energy per one spin projection, while ${\cal E}(T)$ is internal energy of the superconductor with the gap $\Delta$  at temperature $T(x)$ \cite{Kopnin2001}. In Eq.~\equref{ts} we assume that the electronic subsystem releases heat to the normal trap rather than directly to the phonon bath. Indeed, the trap relaxation rates are $\tau_{\rm tr}^{-1}\sim 10^6 ... 10^7$ s$^{-1}$ for the contact resistances $\rho_{\rm tr}\sim 3\ldots 0.3$ k$\Omega\times (\mu m)^2$ while the electron-phonon relaxation rate in aluminium is $\tau^{-1}_{\rm ph}<3\times 10^{5} s^{-1}$ for the experimental temperatures. Equation~\equref{ts} is obtained by averaging it over inhomogeneities in $\Delta$, assumed to have low areal density and short scale compared to the inelastic relaxation length, i.e., the scale of $T(x)$ variations. The thermal conductivity $\kappa$ and the heat current into the trap $q$ are spatially averaged quantities $\kappa =\kappas\left(1-r\right)+\kappan \, r$ and $q=q_{\rm S}\left(1-r\right)+q_{\rm N}\, r$ for a superconductor having a normal fraction $r$ proportional to $\bperp$ which models the presence of vortices. The thermal conductivity~\cite{bardeen59},  $\kappas$, and the heat current, $q_{\rm S}$, of a superconductor at $\kb T\ll\Delta$ are exponentially suppressed relative to their normal-state values $\kappan=\lo\sigman T$ and
$q_{\rm N} =[(\pi^2N_0k_{\rm B}^2/3) (T^2-T_0^2)]/\tau_{\rm tr}$ according to
$\kappas/\kappan=(6/\pi^2)(\Delta/{k_{\rm B}T})^2 e^{-\Delta/k_{\rm B}T}$ and $q_{\rm S} =(2 \pi \Delta)^{3/2}N_0 k_{\rm B}^{1/2}[\sqrt{T}e^{-\Delta/k_{\rm B}T} - \sqrt{T_{0}}e^{-\Delta/k_{\rm B}T_{0}}]/\pi \tau_{\rm tr}$. Here $\lo=(\pi^2/3)(\kb/e)^2$, and $\sigman$ is the Al normal state electrical conductivity.

Because of exponentially small $\kappas$ and $q_{\rm S}$ at $\kb T\ll\Delta$, the temperature $\ts$ can be very sensitive to the vortex fraction $r$. For reference, the black dotted line in Fig.~\ref{fig:zerosmall}~\figtc shows the calculated $\tnmin$ as a function of $\tbath$ in the limit of perfect thermalization, $\ts=T_{\rm ph}=\tbath$. The red and blue solid lines are the results of Eqs.~\equref{nbalance} and~\equref{ts} for equilibrium phonons in the island, $T_{\rm ph}=T_0$. The red line is obtained with $r=0$, so that the S electrode overheats at most all the way to the large-area bonding pad. Especially towards the lowest bath temperatures, the observed cooling in zero field is considerably weaker than the prediction of the model for perfect thermalization of S electrodes. To estimate the effect in the optimum field, we set $r=1$ in the $10\mum$ wide electrode section, but keep $r=0$ in the narrower section. The result is shown as the solid blue line in Fig.~\ref{fig:zerosmall}~\figc. In our samples, a considerable fraction of the N island is located on top of the S electrodes such that $T_{\rm ph}$ can be essentially higher than $T_0$. As a worst-case estimate, we assume the N island phonons to be overheated to $T_{\rm ph}=\ts$. The result for the above two cases of full and partial S electrode overheating are shown as the red and blue dash-dotted lines, respectively. The predicted influence of the field is now stronger, and the measured cooling in both zero and optimum field is bracketed between the solid and the dash-dotted line. Approximate phonon temperatures $T_{\rm ph}$ at  $V_{\rm opt}$ that reproduce exactly the observed $\tnmin$ are displayed in Fig.~\ref{fig:zerosmall}~\fige. In both zero field (red, upper curve) and in the optimum field (lower curve), $\tph$ reflects the temperature $\ts$ of the QPs close to the junction, ranging from close to $0.8\ts$ at $\tbath=0.1\kelvin$ to around $0.95\ts$ at $\tbath=0.5\kelvin$. We stress that the large field-induced improvement evident in Figs.~\ref{fig:scheme} and~\ref{fig:zerosmall} is observed because of the considerable S electrode overheating in zero field. It causes also the significant increase in $\vopt$ as $\bperp$ is increased from zero: $\vopt$ is close to the ideal value $2(\Delta-0.66\kb\tn)/e$~\cite{giazotto06} only at optimum $\bperp$.

\begin{figure}[!htb]
\includegraphics[width=\columnwidth]{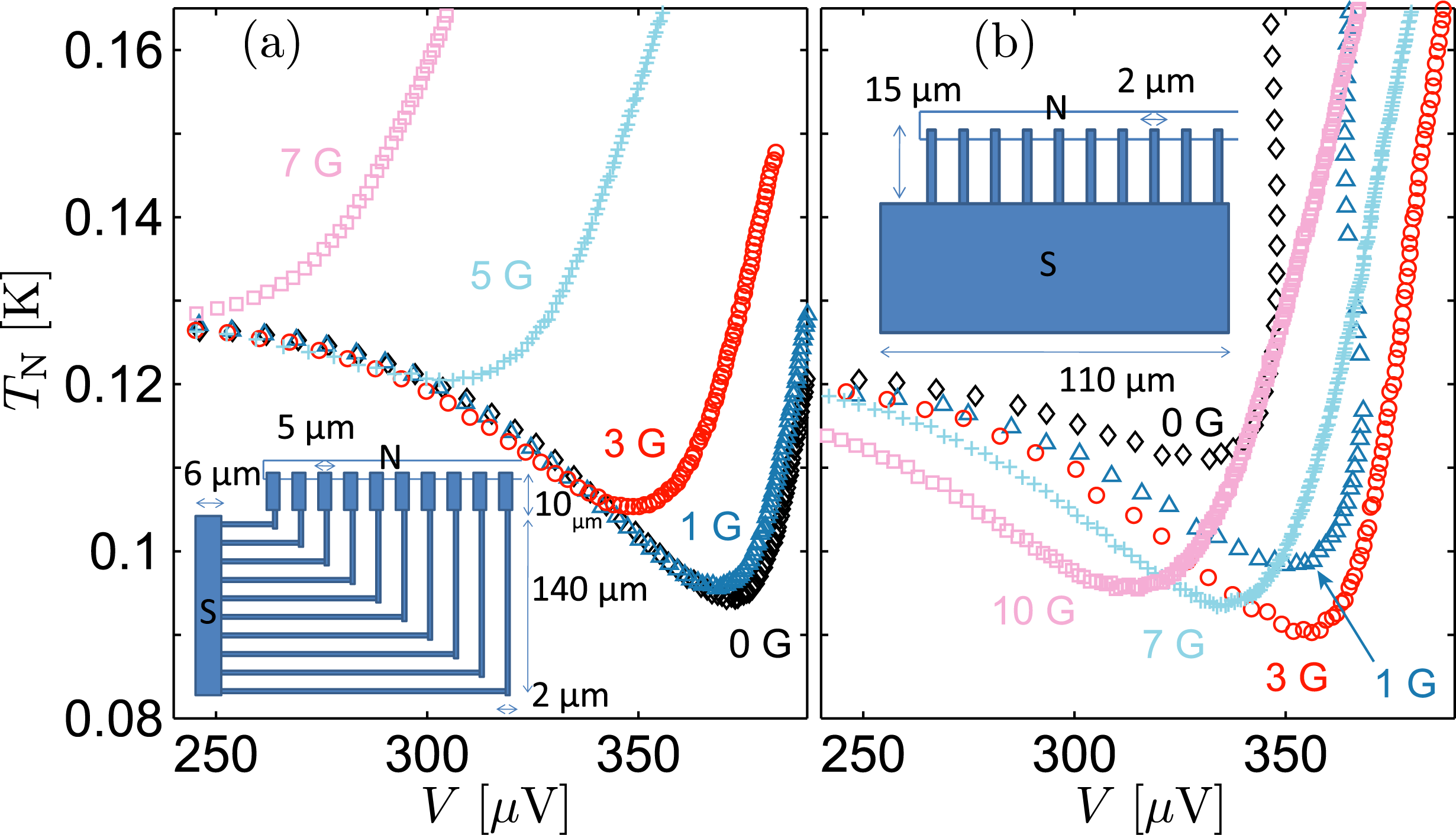}
\caption{(color online) Cooling curves in selected magnetic fields for parallel SINIS coolers with different S electrode geometries, at fixed $\tbath\simeq 130\mk$. \figta When initially wide S leads are followed by a narrower section, applying a finite $\bperp$ weakens the cooling monotonously. \figtb In a structure with initially narrow leads, the behavior is non-monotonous, with enhanced cooling in small fields. The insets sketch one half of the cooler structure with 10 NIS junctions in parallel.} \label{fig:widenarrow}
\end{figure}

To emphasize the role of the S electrode geometry, we performed additional experiments on parallel SINIS coolers. As shown in Fig.~\ref{fig:widenarrow}~\figa, in a sample with initially wide leads, vortices form first close to the junctions, and applying $\bperp$ monotonously weakens the cooling. In contrast, with narrow leads close to the junctions as in Fig.~\ref{fig:widenarrow}~\figb, the cooling is optimized at a finite $\bperp$.

In conclusion, we observed that a small magnetic field enhances relaxation processes in a superconductor thus stabilizing superconductivity in non-equilibrium conditions. Significantly improved electronic cooling in a tunnel junction was achieved in a small perpendicular magnetic field. The enhancement of relaxation can be relevant also for ``re-entrant superconductivity'' observed in Zn nanowires~\cite{chen09,chen11} driven out of equilibrium by supercritical current. A quasiequilibrium model accounts for the field-improved QP relaxation in the S leads. The work can provide means for optimizing the performance of superconducting nanostructures, and sheds additional light on the unsolved problem of nonequilibrium and residual quasiparticles.

\begin{acknowledgments}
We thank H. Courtois, F. Giazotto, F. Hekking and F. Taddei for useful discussions and T. Aref for assistance in sample fabrication. We acknowledge financial support from the European Community's FP7 Programme under Grant Agreement No. 228464 (MICROKELVIN, Capacities Specific Programme), the Academy of Finland (project number 139172), the Finnish Academy of Science and Letters and EPSRC through grant EP/F040784/1. NBK acknowledges support by the Program ``Quantum Physics of Condensed Matter'' of the Russian Academy of Sciences.
\end{acknowledgments}

\end{document}